\documentclass[aps,prb,twocolumn,superscriptaddress,floatfix]{revtex4-2}
\pdfoutput=1 
\usepackage[colorlinks=true,citecolor=blue,linkcolor=blue,breaklinks=true]{hyperref}
\usepackage{amssymb} 
\usepackage{amsmath} 
\usepackage{physics}

\usepackage{times} 
\usepackage{setspace}
\usepackage{xfrac} 
\usepackage{placeins} 

\newcommand{\LCO}{Li$_{2}$CuO$_{2}$ }
\newcommand{\LCOn}{Li$_{2}$CuO$_{2}$}
\begin{document}

%\preprint{}

\title{Refined spin-wave model and multimagnon bound states in \LCO}


\author{Eli Zoghlin}
\affiliation{William H. Miller III Department of Physics and Astronomy, Johns Hopkins University, Baltimore, MD 21218, USA}
\affiliation{Materials Department, University of California, Santa Barbara, CA 93106, USA}

\author{Matthew B. Stone}
\affiliation{Neutron Scattering Division, Oak Ridge National Laboratory, Oak Ridge, TN 37831, USA}

\author{Stephen D. Wilson}
\email[]{stephendwilson@ucsb.edu}
\affiliation{Materials Department, University of California, Santa Barbara, CA 93106, USA}


\date{\today}

\begin{abstract}
Here we report a study of the spin dynamics in the ferromagnetic chain compound \LCOn. Inelastic neutron scattering measurements allow for the spin Hamiltonian to be determined using a $J_1-J_2$ \textit{XXZ}-Heisenberg spin chain model with weak interchain interactions.  The primary exchange parameters determined from our data are qualitatively consistent with those of Lorenz \textit{et al.} [\textit{Europhys. Lett.} \textbf{88}, 37002 (2009)], and our data allow for the resolution of additional interchain exchange interactions. We also observe the formation of two- and, potentially, three-magnon bound states. The two-magnon bound state exists only in the magnetically ordered phase of this material, consistent with stabilization by the weak, Ising-like exchange anisotropy of the nearest-neighbor intrachain interaction. In contrast, the potential three-magnon state persists in a finite temperature regime above $T_{N}$, indicating an unconventional character. Our results establish \LCO  as an experimental platform for the study of exchange anisotropy-stabilized bound states in a ferromagnetic chain.
\end{abstract}

%\keywords{}

\maketitle

\section{Introduction \label{intro}}

Quasi-1-dimensional (quasi-1D) magnetic materials, or ``spin chain'' compounds, are valuable experimental platforms for testing numerous magnetic models due to the analytical simplification that dimensional confinement provides. Furthermore, the suppression of ordering in quasi-1D systems enhances the effects of quantum fluctuations. Accordingly, quasi-1D materials platforms have been found to manifest a range of fascinating phase behaviors, including Haldane singlet formation \cite{schiffler1986, yamashita2000,xu1996,yokoo1995}, fractionalized excitations (e.g., spinons) \cite{nagler1991,tennant1993,tennant1995,kim2006}, quantum criticality \cite{kinross2014,morris2014,faure2018,niesen2013}, and solid state manifestations of quantum few-body phenomena \cite{dally2020,garret1997,tennant2003,bai2021,legros2021,torrance1969,hoogerbeets1984}.

\begin{figure*}
 \includegraphics[width = 17.4 cm]{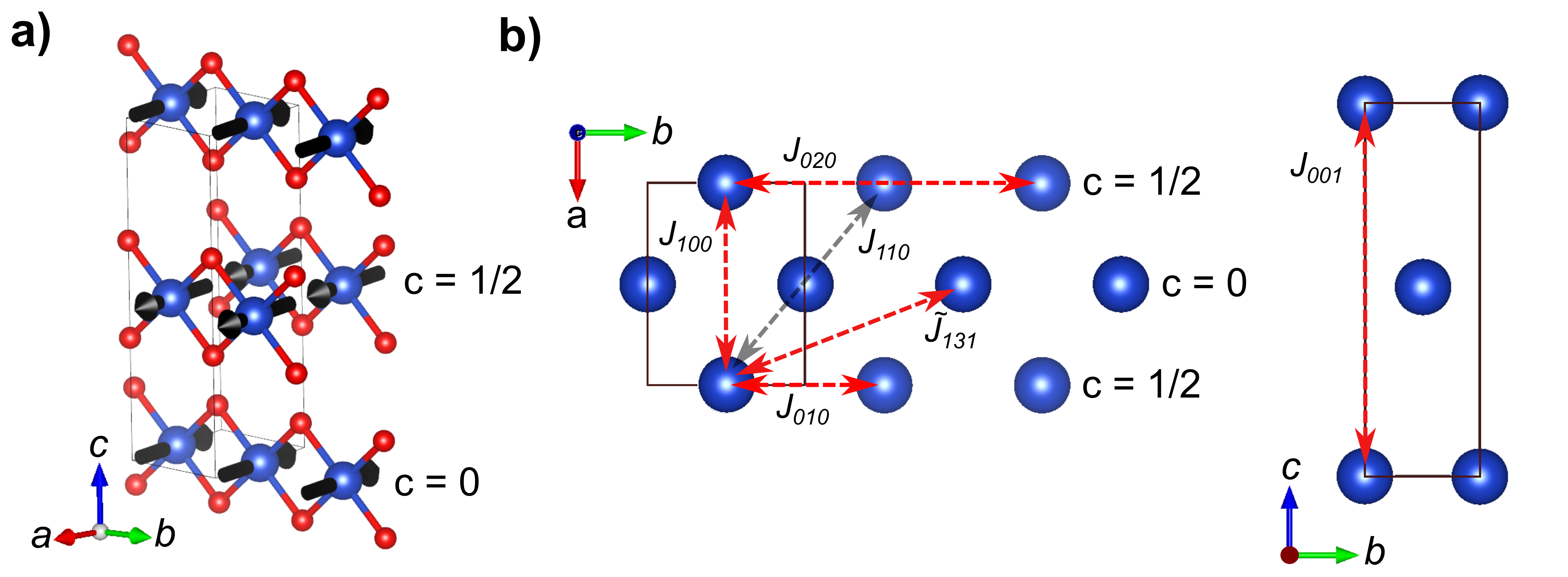}
 \caption{\label{fig:Figure1} \textbf{a)} Atomic crystal (orthorhombic) and magnetic (black arrows) structures of \LCOn, with the shared unit cell shown by the thin grey line (red = O, blue = Cu). The Li atoms have been omitted for clarity. The labels show the $c$ axis coordinate of the two distinct chain subsystems, which are offset from each other in the $ab$ plane by $\frac{1}{2}$($\va*{a} + \va*{b}$). All atomic visualizations were created with VESTA \cite{momma2011}. \textbf{b)} Illustration of the exchange interactions in the $ab$ plane (left) and $bc$ plane (right). The red arrows represent interactions which were included in the final exchange model, while the arrow in black represents the $J_{110}$ interaction which was not included in our final model.}
 \end{figure*}

The magnetic properties of the \textit{S} = $\frac{1}{2}$  spin chain compound \LCO were originally examined more than 30 years ago \cite{sreedhar1988}. The crystal structure (space group \#71, Immm) is composed of chains of edge-sharing, square-planar coordinated Cu$^{2+}$ ions,  prompting later proposals of \LCO as a model compound for exploring magnetism in edge-sharing cuprates \cite{sapina1990, okuda1992, mizuno1998,ebisu1998}. The chains in this system run along the crystallographic \textit{b} axis with each array of chains in the \textit{ab} plane shifted by $\frac{1}{2}$($\va*{a} + \va*{b}$) relative to the neighboring arrays along the \textit{c} axis. An early powder neutron diffraction study reported a commensurate, antiferromagnetic (C-AFM) ground state with magnetic propogation vector $\va*{k}$ = [0 0 1] \cite{sapina1990}, composed of ferromagnetically (FM) aligned chains. Considering an individual array of chains in the \textit{ab} plane, all moments are aligned parallel to the $a$ axis; the moments in the neighboring arrays along the $c$ axis are aligned in the antiparallel direction [see Fig. \ref{fig:Figure1}(a)]. Considerable effort was then made to determine an exchange model that captures the collinear magnetic ground state \cite{yushankhai1999,weht1998,xiang2007,boehm1998,lorenz2009,giri2001}. Despite initial discrepancies in inelastic neutron scattering (INS) results \cite{boehm1998}, the leading interactions in \LCO are now known to be the frustrated FM nearest-neighbor (NN, $J_{010}$) and AFM next-nearest-neighbor (NNN, $J_{020}$) interactions along the chain axis, where the ratio $\alpha$ = $|J_{020}$/$J_{010}|$ defines the degree of frustration \cite{lorenz2009}. This is similar to models of other cuprate spin chain compounds (e.g., LiCuVO$_{4}$ \cite{enderle2005} and PbCuSO$_{4}$(OH)$_{2}$ \cite{wolter2012,rule2017}) which are often described in the context of a Heisenberg $J_{1}-J_{2}$ model (in our notation $J_{1}$ is represented by $J_{010}$ and $J_{2}$ by $J_{020}$).

Of particular interest in the Heisenberg $J_{1}-J_{2}$ model are theoretical phase diagrams \cite{hikihara2008,sudan2009} that indicate the formation of unconventional, multipolar orders in certain regimes of $\alpha$. At small values of $\alpha <$ 1/4, the degree of frustration is sufficiently small that a FM state occurs \cite{bursill1995,bader1979,dimitriev2008}. As $J_{2}$ increases relative to $J_{1}$, multipolar ground states -- which can be thought of as Bose-Einstein-like condensates of \textit{n}-magnon bound states -- are predicted to be stabilized for $\alpha >$ 1/4, with quantum fluctuations preventing the classically predicted long-range ordered spiral state \cite{nishimoto2015, hikihara2008, sudan2009, chubukov1991, syromyatnikov2012}. Within the context of a  $J_{1}-J_{2}$ model, \LCO is a potential candidate for hosting these states ($\alpha \approx$ 1/3 \cite{lorenz2009}); however previous work has also emphasized the importance of interchain interactions (specifically the NNN interaction  $\tilde{J}_{131}$) \cite{lorenz2009,mizuno1998,xiang2007} and exchange anisotropy \cite{dimitriev2009a}. These additional terms suppress fluctuations and are believed to stabilize the observed C-AFM order over the spiral state expected for a $J_{1}-J_{2}$ system with $\alpha >$ 1/4. Nevertheless, the interchain interactions remain reasonably weak compared to the intrachain exchange ($\tilde{J}_{131} \approx$ 0.04$J_{010}$ and 0.1$J_{020}$) \cite{lorenz2009} and, as a result, this material can be considered as an effective system of FM spin chains with the potential to host \textit{n}-magnon bound states.

The existence of multimagnon bound states in FM spin chains, first studied by Bethe in 1931 \cite{bethe1931}, has been the subject of many theoretical studies \cite{dimitriev2009a,dimitriev2009b,haldane1982, schneider1981, southern1994, huang1990, hood1984, kecke2007, onishi2015,sharma2022}. These studies have predicted the existence of a rich array of bound states, with distinct physics arising due to differences in spin and the character of the exchange interactions and anisotropies. CoCl$_{2}\cdotp$2H$_{2}$O provided one of the earliest experimental manifestations of this in the solid state, with far-infrared transmission measurements showing results consistent with \textit{n} $\geq$ 2 bound states \cite{torrance1969}. Indications of a series of \textit{n} $\geq$ 2 bound states have also been seen in (C$_{6}$H$_{11}$NH$_{3}$)CuCl$_{3}$ through FM resonance measurements \cite{hoogerbeets1984}. Suggestively, as in \LCOn, both of these compounds are \textit{S} = $\frac{1}{2}$, two-sublattice AFMs composed of alternating FM chains with easy-axis exchange anisotropy. In fact, the potential for \LCO to host multimagnon bound states has been proposed explicitly in previous theoretical work \cite{dimitriev2009a, dimitriev2009b}. For the integer spin case (\textit{S} = 1), a distinct two-magnon bound state stabilized by single-ion anisotropy has been found in FeI$_{2}$ via far infrared absorption spectroscopy \cite{Fert1978} and INS \cite{petitgrand1979,bai2021}. Subsequent time-domain terahertz spectroscopy measurements have also revealed even higher-order bound states (\textit{n} = 4, 6) \cite{legros2021}. Similar measurements in NiNb$_{2}$O$_{6}$, supported by detailed calculations, also reported evidence of magnon-magnon interactions \cite{chauhan2020,sharma2022}.

Strangely, to the best of our knowledge, the canonical exchange anisotropy-driven magnon bound states of a FM spin chain have yet to be observed by INS \footnote{Note that the exchange model presented in Bai \textit{et al.} for FeI$_{2}$ does include exchange anisotropy. However, the bound state in that case is fundamentally a result of single-ion anisotropy, which provides the energy reduction upon binding. The diagonal exchange anisotropy cooperates with a finite symmetric off-diagonal exchange to hybridize the single-ion anisotropy-stabilized bound state with the one-magnon mode. This explains the visibility of the bound state in INS, as well as the dispersion, but not its stability \cite{bai2021}.}. This represents a substantive gap in the available data since INS, unlike the optical or resonance-based experiments mentioned previously, yields a measure of both the energy \textit{and}  momentum-dependence (away from $|\va*{Q}|$ = 0) of low-energy magnetic excitations. As a result, INS provides additional information about the nature of the magnon-magnon interaction potential.

In this work, we report on exchange anisotropy stabilized two- and, potentially, three-magnon bound states in \LCO observed via INS. Detailed measurements of the one-magnon dispersion are first presented and confirm previous results regarding the importance of the $\tilde{J}_{131}$ interaction in addition to clarifying the nature of further interchain interactions. Here, we also demonstrate explicitly that a uniaxially anisotropic NN intrachain interaction alone fully captures the origin of the gap in the spectrum, i.e., the spin gap. Next, we demonstrate the existence of additional magnetic modes, separate from the one-magnon dispersion, which we ascribe to the formation of two- and, potentially, three-magnon bound states. Analysis of the momentum dependence of the two-magnon state shows a clear splitting along the $H$ and $L$ directions, which most likely stems from the influence of interchain interactions on the momentum dependence of the magnon-magnon interaction potential. These results establish \LCO as an accessible experimental platform for studying the physics of exchange anisotropy-driven bound states in a Heisenberg FM chain via INS and other spectroscopic techniques.

\section{Methods \label{methods}}

The crystal studied in this work was grown using a novel high-pressure floating-zone furnace employing a similar procedure as reported previously \cite{schmehr2019}. In this case, isotopically enriched $^{7}$LiOH$\cdotp$H$_{2}$O (Cambridge Isotopes, 99.9\%) was utilized to mitigate neutron absorption effects. The $^{7}$LiOH$\cdotp$H$_{2}$O was dried at 150 \textdegree C under a rough, dynamic vacuum for 12 h resulting in anhydrous $^{7}$LiOH. Polycrystalline powder was then produced by grinding together a stoichiometric amount of CuO (Fisher, 99.999\%) with $^{7}$LiOH, using 5 mass \% excess of $^{7}$LiOH over the stoichiometric amount. All grinding was carried out in an inert atmosphere glove box. The difference in the required excess of LiOH -- determined by the amount needed for full reaction of the CuO -- relative to our previous report \cite{schmehr2019} is due to differences in the water content of the LiOH. The well-mixed powder was then pressed into a pellet at 300 MPa using a cold isostatic press and fired at 750 \textdegree C in air for 24 h. The reacted powder was then pressed into seed and feed rods ($\approx$ 4 mm diameter) at 300 MPa and fired in air for an additional 32 h at 750 \textdegree C on a sacrificial bed of (isotopically enriched) \LCO powder. Growth was conducted under 100 bar of an Ar:O$_{2}$ = 80:20 mixture. The feed and seed rods were counter rotated at 8 and 9 rpm, respectively, and a growth rate of 10 mm/h was used \cite{wizent2011}.

 \begin{figure}
 \includegraphics[width = 8.6 cm]{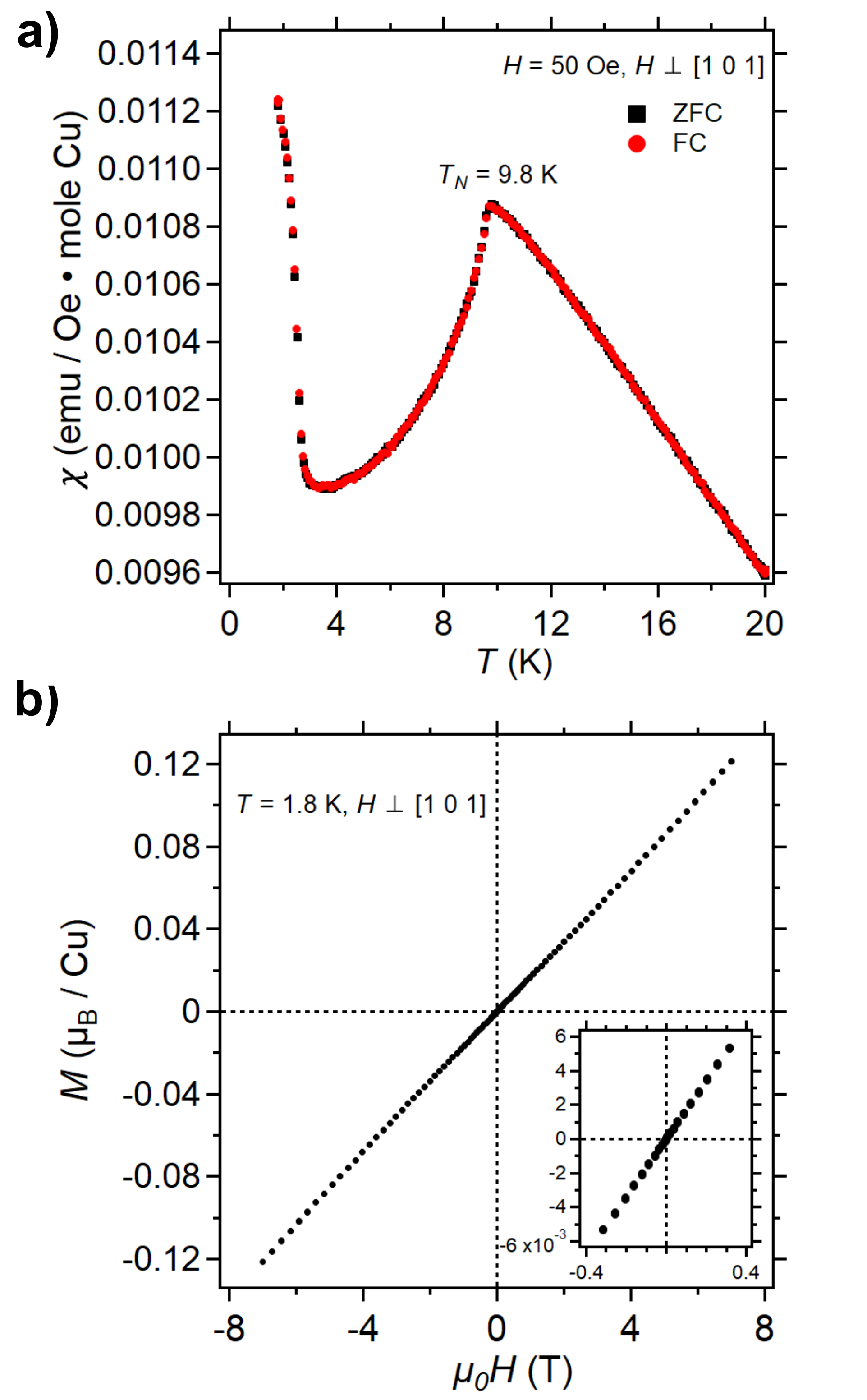}
 \caption{\label{fig:Figure2}\textbf{ a)} Magnetic susceptibility ($\chi \equiv M$/$H$) at low temperature showing $T_{N}$ = 9.8 K. Note the lack of irreversibility at low-temperature ($T <$ 3 K). \textbf{b)} Field dependence of the magnetization ($M$) at low-temperature. \textit{Inset:} Low-field ($\mu_{0}H <$ 0.5 T) region.}
 \end{figure}
 
The phase purity and lattice parameters of the resulting crystal were characterized using laboratory powder X-ray diffraction (Emperyan, Panalytical). Powder diffraction data were collected on crushed crystal pieces and showed a nearly phase pure sample with the presence of a very small amount (estimated to be $<$ 1 mass \%) of an unidentified impurity. Comparison of measurements conducted on powders produced from crystalline material collected near the boule's surface versus the boule's core suggests this impurity occurs as a very thin surface layer. The refined lattice parameters were: $a$ = 3.66171(2) \r{A}, $b$ = 2.86436(7) \r{A}, and $c$ = 9.3945(2) \r{A}, consistent with literature values \cite{sapina1990}. 

Bulk magnetic properties were measured on a small crystal ($\approx$ 10 mg) cleaved from the main boule, which was mounted on a quartz paddle using GE varnish such that the applied field lay within the (1 0 1) plane. Measurements were conducted using a Quantum Design MPMS3 SQUID magnetometer in VSM mode. Field-cooled and zero-field-cooled magnetic susceptibility measurements [Fig. \ref{fig:Figure2}(a)] show $T_{N}$ = 9.8 K, with no sign of hysteresis. The field-dependence of the magnetization at $T$ = 1.8 K [Fig. \ref{fig:Figure2}(b)] is linear across the full measured range ($\mu_0 H = \pm$ 7 T), with no observable hysteresis. Previous chemical analysis on a crystal grown in a very similar manner showed Li:Cu = 1.99:1.01 \cite{schmehr2019}, and the lack of any low-temperature ferromagnetism suggests a fully occupied oxygen site \cite{shu2018}. The measured $T_{N}$ is elevated compared to previous results for crystals grown at lower pressures ($\approx$ 8 $-$ 9 K) \cite{chung2003,lorenz2009,shu2018,kawamata2004}, likely reflecting an optimized stoichiometry. The upturn at $T < 3$ K appears to be unrelated to the small, unidentified, extrinsic impurity in our sample since it also occurs in reportedly phase-pure samples \cite{shu2018}. It is most likely the result of a small volume fraction of intrinsic paramagnetic impurity spins stemming from remnant defects.

\LCO cleaves easily along the (1 0 1) plane and the crystal separated into pieces upon removal from the furnace. Due to the brittle nature of the fracture, the pieces could be reassembled using Al wire with reasonably good registry. The resulting sample ($\approx$ 2 g) was attached to an Al mount and aligned in the (0 \textit{K L}) scattering plane using the two-axis alignment station at the High-Flux Isotope Reactor (HFIR). INS measurements were then conducted using the direct geometry time-of-flight (TOF) spectrometer SEQUOIA at the Spallation Neutron Source (SNS). The aligned sample and mount were sealed under 1 atm of He exchange gas inside an Al can to allow for temperature control. A thermocouple was mounted to the sample can which was then attached to the cold finger of a bottom loading closed-cycle refrigerator (base temperature $\approx$ 5 K) on SEQUOIA. The available beam size at SEQUOIA is 50 $\times$ 50 mm, which can be adjusted by a motorized aperture whose horizontal and vertical dimensions can be altered independently. The aperture was centered on the sample and its opening size was adjusted such that the incident beam profile matched the size of the sample. For background measurements the aperture was translated from the initial position so as to no longer illuminate the sample, but to continue to illuminate the sample can and its helium exchange gas. These measurement were performed in the identical instrument configurations as the sample measurement, but for only a single orientation of the sample rotation axis. Independent measurements were taken with vertical and horizontal translations of the aperture; these were found to be the same and therefore combined. To produce a suitably smooth data set for background subtraction in the line cuts the ``pseudo-empty'' can data was integrated in the proper $|\va*{Q}|$-range and then fit to an exponential decay function, $I(\Delta E)$ = $y_{0}$ + Aexp(-$\tau\Delta E$). This function yielded an essentially flat background in the main $\Delta E$-range of interest where it is most relevant (2 -- 5 meV).

\begin{figure}
 \includegraphics[width = 8.6 cm]{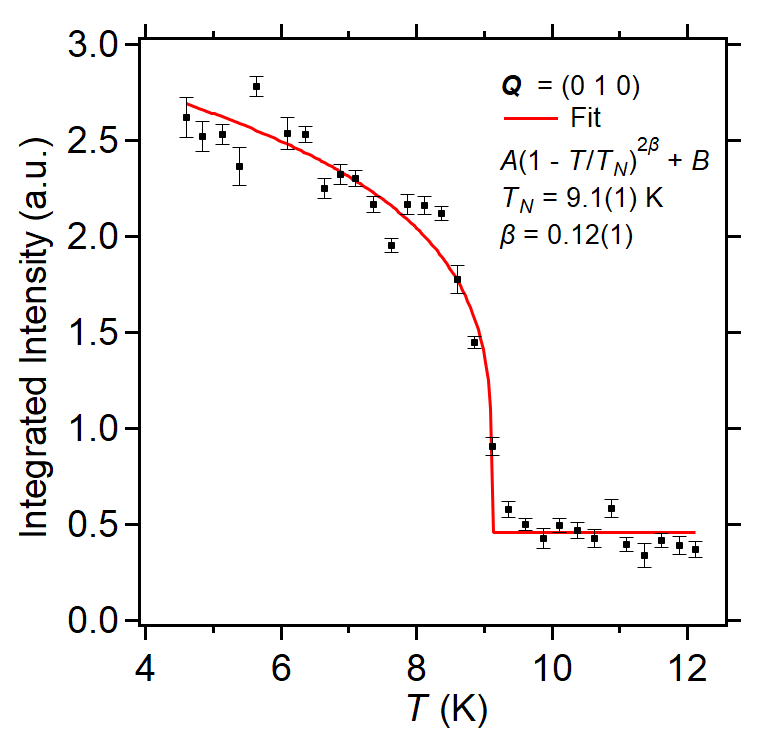}
 \caption{\label{fig:Figure3} Order parameter tracking the integrated intensity of the $\va*{Q}$ = (0 1 0) magnetic Bragg peak across the C-AFM phase transition. The error bars reflect the propogated uncertainties from the intensity data. The red line is the indicated fit to the data, described further in the main text.}
 \end{figure}

Measurements were conducted by rotating the crystal around the vertical axis of the instrument in steps of 1$^\circ$ or 2$^\circ$. Several independent measurements were conducted using a range of incident energies ($E_{i}$ = 11, 25, 50 and 185 meV with each measurement corresponding to a single $E_{i}$) in order to access the full bandwidth of the highly dispersive intrachain excitations. The data were reduced using the software package Mantid \cite{arnold2014}. Where possible, the data were symmetrized by folding in order to improve the counting statistics. All error bars in intensities represent one standard deviation. An order parameter was collected by fixing the rotation angle of the crystal such that the $\va*{Q}$ = (0 1 0) magnetic Bragg peak could be measured and then sweeping the temperature at $\approx$ 0.5 K/min. The data were binned in temperature with a bin width of 0.25 K and powder integrated to give intensity as a function of $|\va*{Q}|$ for elastically scattered neutrons. The integrated peak intensity was extracted by numerically integrating the data at each temperature within a consistent $|\va*{Q}|$-range. The resulting order parameter was fit (red line, Fig. \ref{fig:Figure3}) with the following function: $A(1 - T/T_{N})^{2\beta} + B$, where $A$ is the integrated magnetic scattering intensity in the fully ordered state, $\beta$ is the associated critical exponent, and $B$ is the background integrated scattering intensity. This analysis gave $T_{N}$ = 9.1(1) K and $\beta$ = 0.12(1).

The dispersion curves along high-symmetry directions were initially extracted using Gaussian fits to constant-$\va*{Q}$ (wave-vector transfer) and constant-$\Delta E$ (energy transfer) cuts through the data; constant-$\Delta E$ cuts were only used for the steep portions of the dispersion along $K$ (the chain direction). Analysis of the extracted dispersion was then conducted using the SpinW software package \cite{toth2015}. For validation, the data were also directly fit using the analytical form of the dispersion relation from linear spin wave theory; both approaches gave very similar results. Magnetic structure optimization was also performed with SpinW. The optimal propagation vector was determined by the Luttinger-Tisza method and the moment direction was optimized to obtain the lowest energy configuration consistent with that propagation vector. 

\section{Spin Wave Measurements and Modeling \label{results}}

\begin{figure*}[h!]
 \includegraphics[width = 17.4 cm]{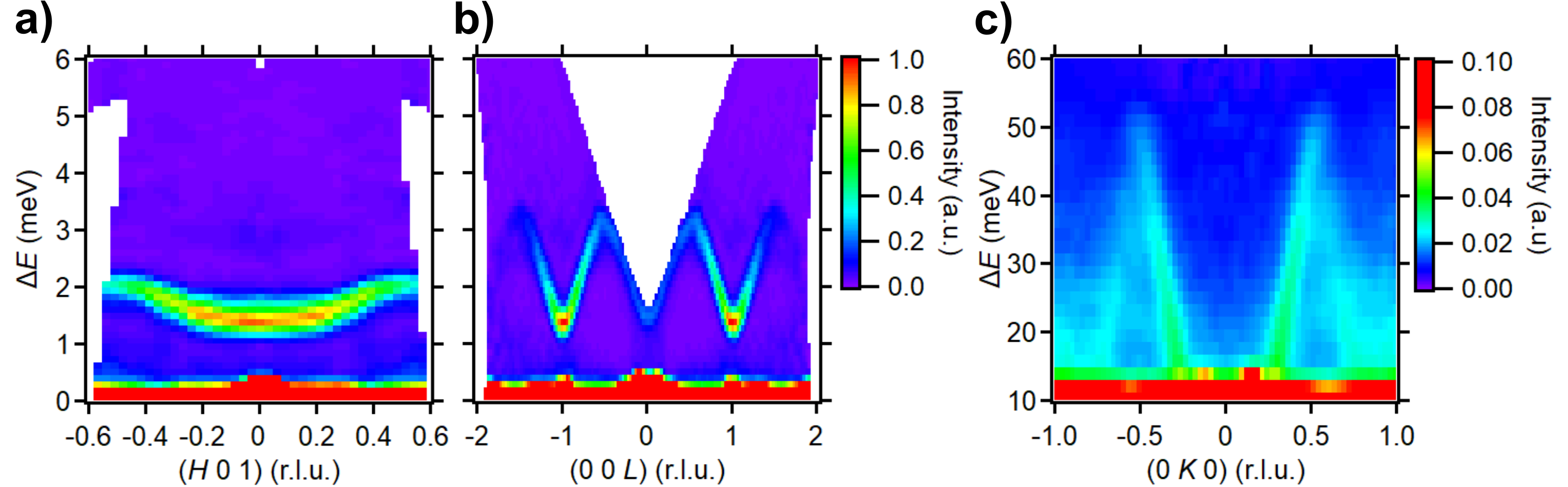}
 \caption{\label{fig:Figure4} Neutron scattering data collected at $T$ = 5 K along high-symmetry directions. The same scale is used in all plots. \textbf{a)} Scattering intensity along the ($H$ 0 1) direction, collected using $E_{i}$ = 11 meV. The integration bounds are: $K$ = [-0.025, 0.025] and $L$ = [0.875, 1.125]. \textbf{b)} Scattering intensity along the (0 0 $L$) direction, collected using $E_{i}$ = 11 meV. The integration bounds are: $H$ = [-0.125, 0.125] and $K$ = [-0.075, 0.075]. \textbf{c)} Scattering intensity along (0 $K$ 0) direction collected out to the zone boundary $\Delta E$ using $E_{i}$ = 185 meV. The integration bounds are: $H$ = [-2, 2] and $L$ = [-2, 2].}
 \end{figure*}

INS data were collected using multiple $E_{i}$'s in the ordered state ($T=5$ K), allowing the spin waves along almost all high-symmetry directions in the Brillouin zone to be resolved. In Fig. \ref{fig:Figure4}(a) and (b), $E_{i}$ = 11 meV data show the magnon dispersion orthogonal to the chain direction for $\Delta E <$ 6 meV. Consistent with previous measurements, both interchain directions are weakly dispersive relative to the intrachain dispersion \cite{boehm1998,lorenz2009}. In Fig. \ref{fig:Figure4}(c), the much steeper intrachain dispersion is revealed by the $E_{i}$ = 185 meV data, which approaches a zone boundary energy of $\approx$ 40 meV.  As discussed below, the dispersion could be well parameterized by linear spin wave theory.

The spin system of \LCO has previously been described \cite{boehm1998,lorenz2009} by the following Hamiltonian:  

\begin{equation}
\hat{H}=\frac{1}{2} \sum_{\mathbf{n}, \mathbf{r}}\left[J_{\mathbf{r}}^{z} \hat{S}_{\mathbf{n}}^{z} \hat{S}_{\mathbf{n}+\mathbf{r}}^{z}+J_{\mathbf{r}} \hat{S}_{\mathbf{n}}^{+} \hat{S}_{\mathbf{n}+\mathbf{r}}^{-}\right]
\end{equation} 

Here, $J_{\mathbf{r}}^{z}$ is the component of the exchange interaction along the $a$ axis (moment direction), determined to be the magnetic easy-axis by electron spin resonance measurements \cite{ohta1993, kawamata2004}. The double sum is over the Cu-sites ($\mathbf{n}$) and the lattice vectors ($\mathbf{r}$) which connect Cu-sites. Using the convention that $J_{r} <$ 0 is FM, the dispersion relation derived via linear spin wave theory \cite{oguchi1960} for this Hamiltonian is given by the following relation:

\begin{equation}
\omega_{\mathbf{Q}}=\sqrt{\left(J_{\mathbf{Q}}-J_{\mathbf{0}}+\tilde{J}_{\mathbf{0}}-D\right)^{2}-\left(\tilde{J}_{\mathbf{Q}}\right)^{2}},
\end{equation}with $J_{Q} \equiv (1/2) \sum_{\mathbf{r}}J_{r}exp(i\va*{Q}\cdotp\va*{r})$ and  $J_{0} \equiv (1/2)\sum_{r}J_{r}$ (the analogous definitions apply to $\tilde{J}$). The $\va*{k}$ = [0 0 1] magnetic structure breaks the body-centered symmetry of the nuclear unit cell, creating two inequivalent Cu-sites and therefore two FM chain subsystems. The $J$ parameters represent intra- and interchain interactions within a subsystem, while the $\tilde{J}$ parameters represent interchain interactions which connect the two subsystems. In this analytical form of the dispersion, $D$ parameterizes the magnitude of the exchange anisotropy according to: $D \equiv J_{\mathbf{0}}^{z}  - J_{0} - \tilde{J}_{0}^{z} + \tilde{J}_{0}$. The spin gap, $\Delta E_{N=1}$, arises due to the finite $D$ but is also influenced by the magnitude of the inter-sublattice interactions: 

\begin{equation}
\Delta E_{N=1} = \sqrt{D(D - 2\tilde{J}_{0})} .
\end{equation} Within the SpinW implementation of linear spin wave theory the individual components of the exchange matrices are fitting parameters.

\begin{figure*}[h!]
 \includegraphics[width = 17.4 cm]{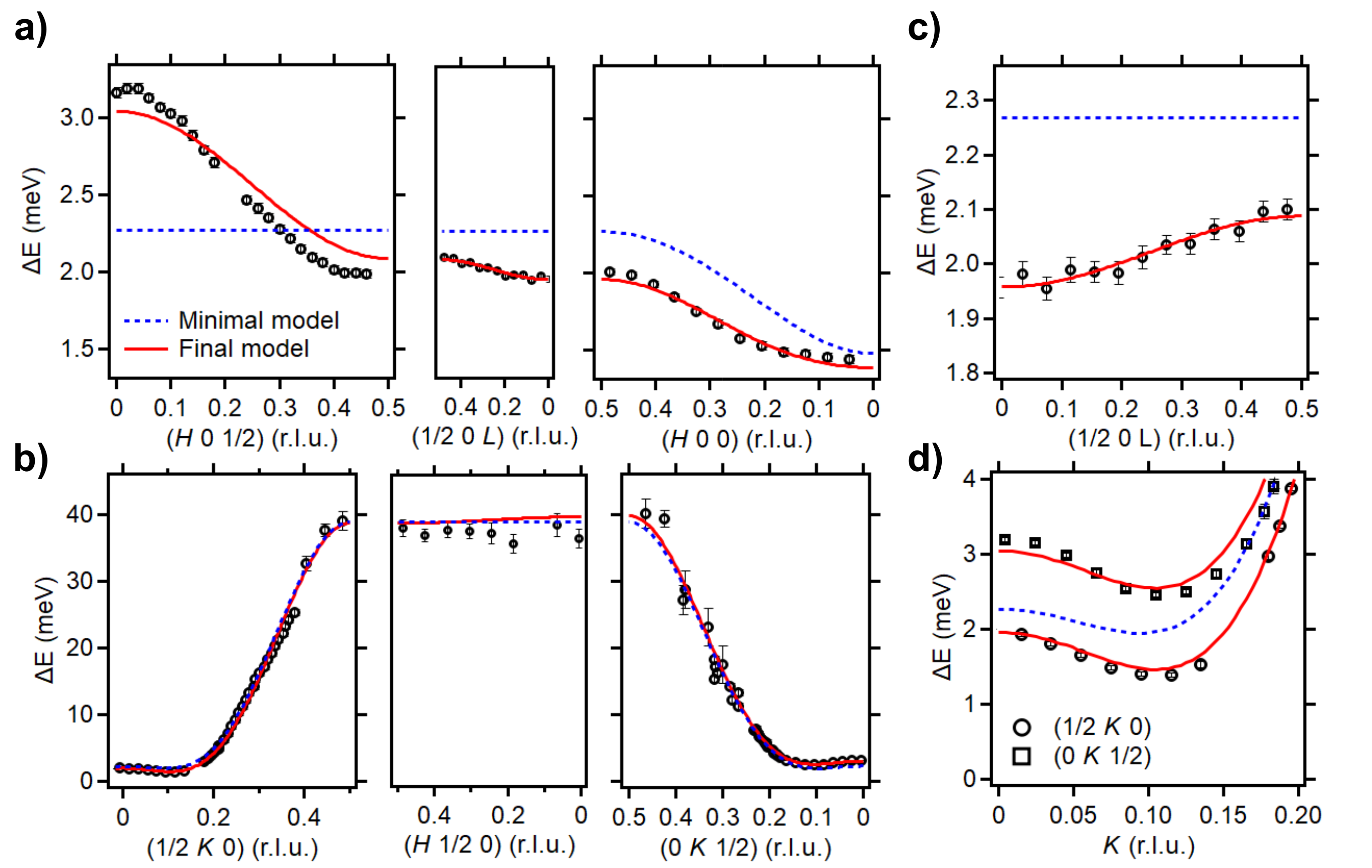}
 \caption{\label{fig:Figure5}\textbf{a)} - \textbf{b)} The extracted one-magnon (N = 1) dispersion along several high-symmetry directions; the $x$-axes are scaled by the magnitude of the corresponding reciprocal space basis vector ($|$$\va*{c}$*$|$ $< |\va*{a}$*$|$ $< |\va*{b}$*$|$). The solid red line shows the fit resulting from the final model described in the text. The dashed blue line shows the fit resulting from the minimal model without $J_{100}$, $J_{110}$ or $J_{001}$. \textbf{c)} Closer view of the (1/2 0 $L$) dispersion showing the small, but finite, bandwidth.\textbf{ d)} Closer view of the difference between the (1/2 $K$ 0) and (0 $K$ 1/2) dispersions near the $X$- and $Z$-points, respectively.}
 \end{figure*}

The extracted dispersion relations from the data shown in part in Fig. \ref{fig:Figure4} are plotted along select directions in Fig. \ref{fig:Figure5}. These were initially fit (blue dashed line in Fig. \ref{fig:Figure5}) using the previously reported minimal model \cite{lorenz2009} that includes only the $J_{010}$, $J_{020}$ and $\tilde{J}_{131}$ exchange interactions as well as exchange anisotropy (see Fig. \ref{fig:Figure1}(b) for an illustration of all considered exchange interactions -- note that $\tilde{J}_{131}$ is associated with $\va*{r}$ = (1/2)[1 3 1]). $J_{010}$ was set to have an easy-axis anisotropy  \cite{sapina1990} (i.e., $J_{010}^{z}$ was fit independently of $J_{010}^{x} = J_{010}^{y} = J_{010}$) since it is by far the largest energy scale. This choice is consistent with theoretical expectations given the nature of a 90$^\circ$ Cu-O-Cu superexchange pathway (the angle in \LCO is $\approx$ 94$^\circ$) \cite{yushankhai1999}. All other exchange interactions were then set to be isotropic. As shown in Fig. \ref{fig:Figure5}, while this minimal model largely captures the behavior of the dispersion within the (0 $K$ $L$) plane, including the magnitude of the spin gap, there are a number of qualitative failures in capturing the dispersion along trajectories with a finite $H$-component. 

The first qualitative failure of the minimal model is the prediction of a flat mode along the ($H$ 0 1/2) direction, where the data clearly show a finite bandwidth [Fig. \ref{fig:Figure5}(a)]. A second failure appears in comparing the dispersions along the (1/2 $K$ 0) and (0 $K$ 1/2) directions. While the minimal model gives an identical dispersion relation for these two directions, the data show a clear difference between them. As shown in Fig. \ref{fig:Figure5}(d) this difference is most apparent in the regions near the $X$- and $Z$-points (i.e., the region approaching $K$ = 0 in the plot). The third qualitative failure is more subtle. The data show a small, but finite, bandwidth to the dispersion along the (1/2 0 $L$) direction, seen most clearly in Fig. \ref{fig:Figure5}(c), which is not captured within the minimal model. 

These departures from the minimal model motivate the inclusion of additional exchange interactions to provide a Hamiltonian which fully represents the measured dispersions. The analytical form of the dispersion shows that only an intra-sublattice exchange with a component along $L$ can induce finite dispersion along the (1/2 0 $L$) direction. Given the relatively large $c$ axis lattice parameter (9.3945(2) \r{A}), $J_{001}$ (with the smallest associated real space distance) is the most reasonable exchange interaction to reproduce this feature. To address the departures observed in the ($H$ 0 1/2), (0 $K$ 1/2), and (1/2 $K$ 0) dispersion relations, additional exchange terms with components along $H$ are required. This suggests the potential inclusion of two additional terms, $J_{110}$ and $J_{100}$. Fitting the dispersion with the inclusion of all three of these exchange interactions greatly improves the fits to the data. In this modified model, $J_{100}$ was found to be AFM and of a similar magnitude as $\tilde{J}_{131}$, which was found to be $\approx$ 7.7 K. $J_{110}$ and $J_{001}$ were both found to be an order of magnitude smaller and FM. 

The modified model with both $J_{110}$ and $J_{100}$ terms significantly decreased the $\chi^{2}$ value of the refinement. However, the inclusion of both terms was found to not be of equal significance. If $J_{100}$ was removed from the model the $\chi^{2}$ value increased by 37\%, while for the model without $J_{110}$ the $\chi^{2}$ value increased by only 1.5\%. The latter increase is marginal and simply due to the decrease in the total number of fit parameters (from seven to six). Therefore $J_{110}$ was set to zero in the final fits. Figure \ref{fig:Figure5} shows the resulting dispersion relations from this final model (solid red line) overplotted with the data and the corresponding parameters are presented in Table \ref{tab:Table1}. We note here that, despite the significance of the interchain interactions with respect to the the magnetic properties, including the nature of the long-range order \cite{ lorenz2009, mizuno1998, giri2001,xiang2007} and the largely undiminished Cu$^{2+}$ moment (0.96(4) $\mu_{B}$) \cite{sapina1990}, the intrachain interactions are still clearly the dominant energy scale. 

 \begin{table*}[]
 \caption{\label{tab:Table1}Comparison of the final exchange parameters (given in Kelvin) determined in this work with corresponding values given in Lorenz \textit{et al.} \cite{lorenz2009} (the ``minimal'' mode). Values in italics were fixed to zero. In our results, $J_{010}^{z}$ (-217.4(9) K ) and $J_{010}^{y}  = J_{010}^{x}  = J_{010}$ are separate fit parameters. As a result,  $D$ is no longer a fitting parameter but is calculated as $D=J_{010}^{z} - J_{010}$. Since $D$ is a small difference between two large values, the resulting error is rather large -- direct fits to the analytical form of the dispersion gave $D$ with a lower error bar, and a value more similar to the results from Lorenz \textit{et al.} }
 \begin{ruledtabular}
\begin{tabular*}{17cm}{c|c|c|c|c|c|c|c|c} 
  & $J_{010}$ & $J_{020}$ & $\alpha =|J_{020} / J_{010}|$ & $\tilde{J}_{131}$ & $J_{100}$ & $J_{110}$ & $J_{001}$ & $D$ \\ \hline
  & & & & & & & &  \\
  This work & -213.3(9) & 67.2(3) & 0.315(2) & 7.44(2) & 5.55(4) & \textit{0} & -0.75(4) & -4(1) \\ 
  & & & & & & & &  \\
  Lorenz \textit{et al.} & -228(5) & 76(2) & 0.332(5) & 9.04(5) & \textit{0} & \textit{0} & \textit{0} & -3.3(2) \\
 \end{tabular*}
 \end{ruledtabular}
 \end{table*}

\section{multimagnon Bound States}

\begin{figure*}[ht]
 \includegraphics[width = 17.4 cm]{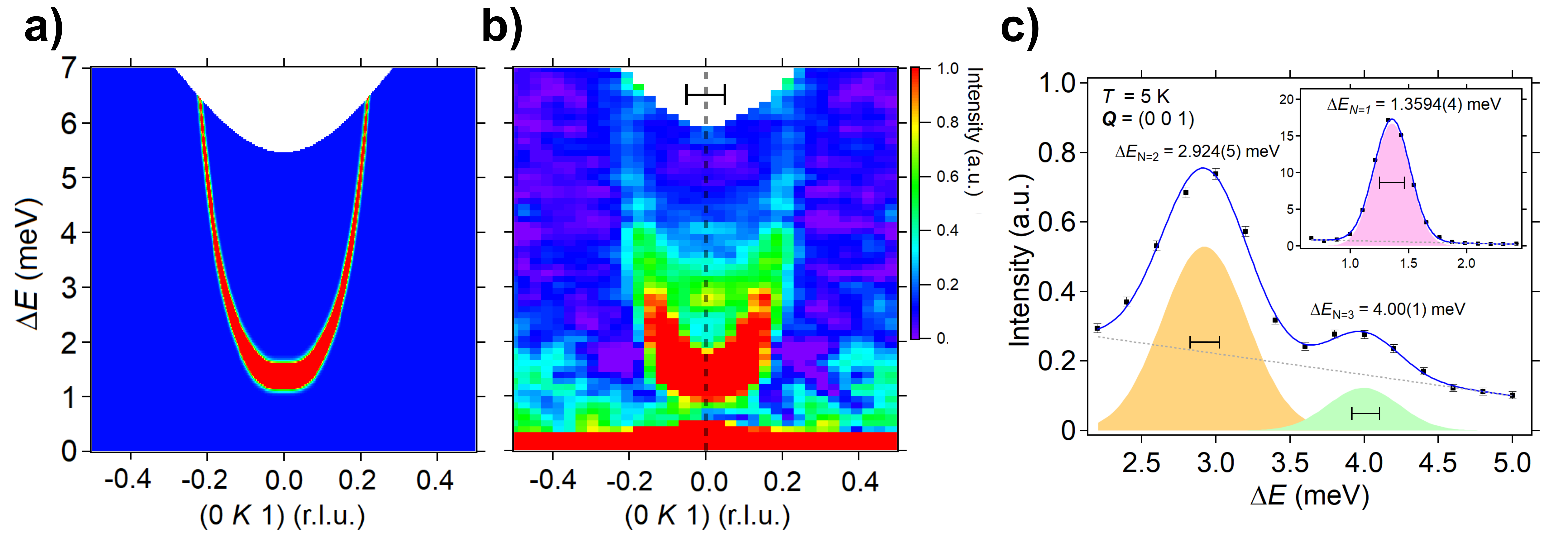}
 \caption{\label{fig:Figure6}\textbf{a)} Spin wave calculation of the $T$ = 0 K, one-magnon dispersion along the (0 $K$ 1) direction  using the parameters in Table I. The calculation includes convolution with the $\Delta E$-dependent resolution function and the $\va*{Q}$-resolution was set to the Gaussian full-width-at-half-maximum (FWHM) of the (0 0 1) magnetic Bragg peak. A small constant background term was added to simulate the incoherent background in the measured data.  \textbf{b)} Measured INS intensities for the same region of  ($\va*{Q}$, $\Delta E$) at $T$ = 5 K using $E_{i}$ = 11 meV. The data were integrated with the following bounds: $H$ = [-0.125, 0.125], $L$ = [0.95, 1.05]. The pseudo-empty Al data has been subtracted directly. Some broad background features remain at $\Delta E <$ 2 meV and any pixels of slight over subtraction have been set to have zero intensity. In \textbf{a} and \textbf{b} the shared color scaling was set by first normalizing the two data sets to the value at $K$ = 0 and $\Delta E$ = $\Delta E_{1}$. To highlight the weak, higher-energy features observed at the $\Gamma$-point in \textbf{b}, the maximum of the color scaling was set to a low value. Both data sets were then normalized once more to this new maximum value. \textbf{c)} $\Delta E$ cut along the dashed line in \textbf{b} (the bar at the top shows the $K$-integration region), with the Al background subtracted as described in the text, showing the modes above the one-magnon dispersion (N = 2 and 3). The data has been scaled by the same overall factor as in \textbf{b}. The blue line is a fit using Gaussians on a linear background; the solid color regions show the Gaussian components and the grey dashed line shows the linear background. The N = 2 and 3 modes are fit together as a sum and the N = 1 mode is fit separately using a higher $\Delta E$ point density. The black lines show the instrumental $\Delta E$-resolution (FWHM). The cut was integrated in reciprocal space with the following bounds: $H$ = [-0.075, 0.075], $K$ = [-0.05, 0.05], and $L$ = [0.95, 1.05]. \textit{Inset:} The one-magnon (N = 1) peak using the same reciprocal space integration but a higher $\Delta E$ point density.}
 \end{figure*}

In addition to resolving the one-magnon dispersion relation, low-energy INS data also reveal additional modes near the magnetic zone-center.  As a baseline illustrating the expectations of one-magnon scattering, Fig. \ref{fig:Figure6}(a) shows a select ($K$, $\Delta E$)-slice populated with the single-magnon spectral weight and dispersion trajectories calculated by SpinW from our exchange parameters and anisotropy values in Table \ref{tab:Table1}.  INS data collected using $E_{i}$ = 11 meV at $T$ = 5 K show the scattering in this same ($K$, $\Delta E$)-slice in Fig. \ref{fig:Figure6} (b). Additional spectral weight appears in the data above the one-magnon gap at $\Delta E_{N=1}$ and, to further parameterize this, a constant momentum cut is plotted in Fig. \ref{fig:Figure6}(c) along the dashed line shown in Fig. \ref{fig:Figure6}(b). This cut along $\Delta E$ through the $K=0$ magnetic zone center reveals two additional features above $\Delta E_{N=1}$: a second peak appears at $\Delta E_{N=2}\approx 2\times\Delta E_{N=1}$, and a third, weaker, peak appears at $\Delta E_{N=3}\approx 3\times\Delta E_{N=1}$. The feature at $\Delta E_{N=2}$ is also clearly visible in the slices presented in Fig. \ref{fig:Figure4}(a) and (b), provided the intensity is set to a lower maximum value that saturates the one-magnon signal. As a further check against the expectations of one-magnon scattering, we have attempted to reproduce these features in our SpinW modeling by arbitrarily setting the diagonal components of $J_{010}$ to have the full $XYZ$-type anistropy allowed by the orthorhombic symmetry. The degeneracy of the one-magnon mode can be broken in this way, but the overall change to the dispersion is inconsistent with our observations. Furthermore, this mechanism cannot create the third mode at $\Delta E_{N=3}$.

An additional, even weaker feature is visible near $\Delta E \approx$ 5.5 meV in the heat map; however, it appears near the edge of the accesible $\Delta E$-range, making analysis less reliable. Furthermore, this region is partially polluted by an optical phonon at slightly higher energies, making it difficult to unambiguously isolate any magnetic signal. We defer further exploration of this potential $\Delta E_{N=4}$ mode to future experiments.

\begin{figure}[h]
 \includegraphics[width = 8.6 cm]{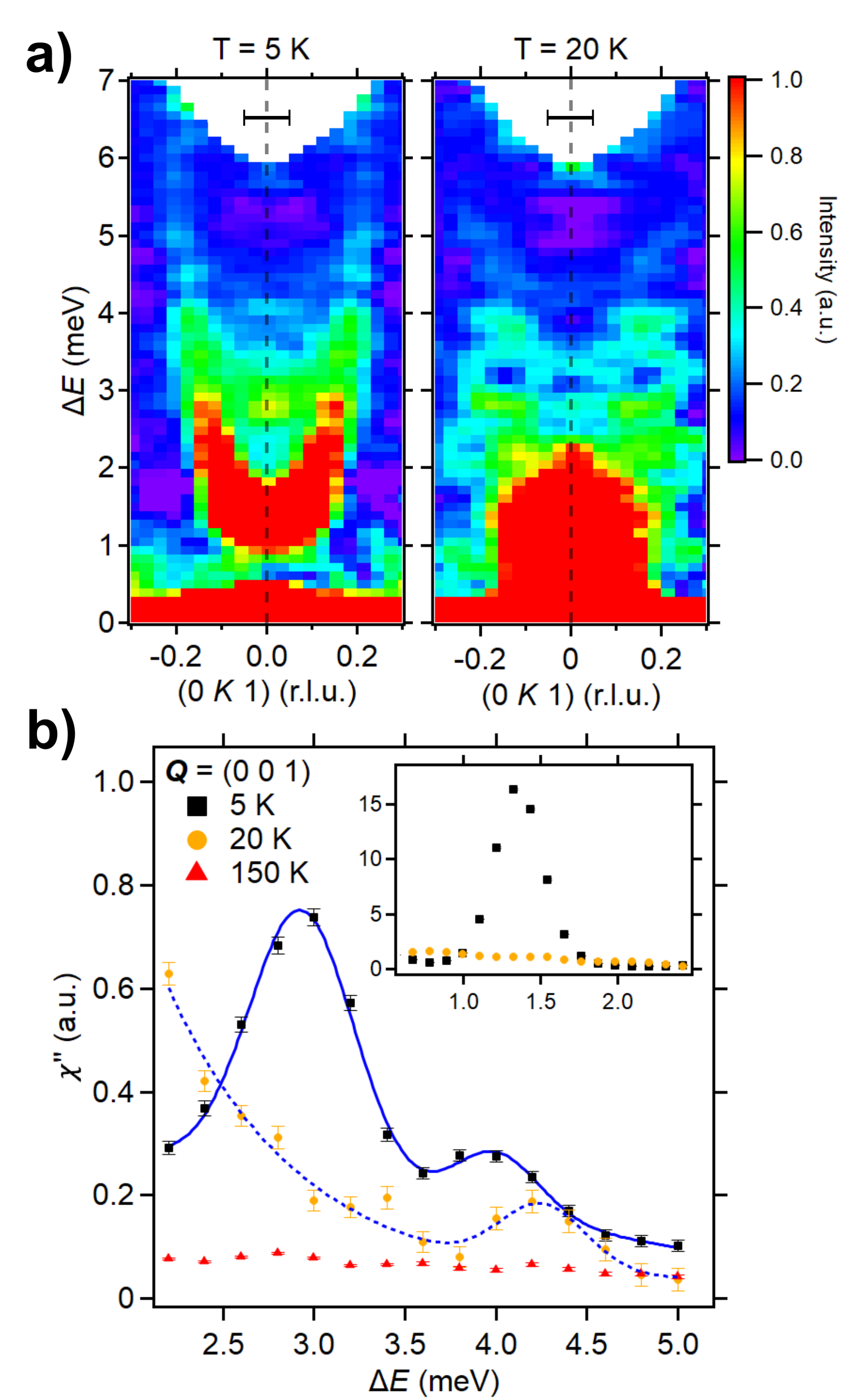}
 \caption{\label{fig:Figure7}\textbf{a)} Temperature dependence of the INS intensities along the (0 $K$ 1) direction collected using $E_{1}$  = 11 meV. The integration ranges are: $H$ = [-0.125, 0.125], $L$ = [0.95,1.05]. The Al background has been subtracted. \textbf{b)} Comparison of the same energy cut shown in Fig. \ref{fig:Figure6}(c) for  $T$ = 5 K ($T< T_{N}$), 20 K ($T> T_{N}$) and 150 K ($T >> T_{N}$). The data has been converted to the dynamic susceptibility using the detailed balance correction, $\chi''$($\va*{Q}$, $\Delta E$) = $\pi(1 - e^{-\Delta E/k_{B}T})S$($\va*{Q}$, $\Delta E$), after subtracting the Al background. The data was then scaled using the same factor as in Fig. \ref{fig:Figure6}. The blue lines are fits using Gaussians plus a linear (solid) or exponential decay (dashed) background. The Gaussian fit for the N = 3 excitation at 20 K uses the FWHM resulting from the fit at 5 K. \textit{Inset:} View of the one-magnon peak region at 5 K and 20 K, also plotted as $\chi''$, with a higher $\Delta E$ point density }
 \end{figure}

A comparison of the (0 $K$ 1) slices collected at $T$ = 5 K and $T$ = 20 K is shown in Fig. \ref{fig:Figure7}(a). This comparison reveals the collapse of the one-magnon gap upon heating out of the ordered state, consistent with its origin from weak exchange anisotropy (magnitude $\approx$ 4 K). Further demonstration of this is given by the zone center energy cuts shown in the inset of Fig. \ref{fig:Figure7}(b). The analogous cut in the main panel of this figure also shows that strong paramagnon scattering remains at 20 K, as is illustrated by the slice in Fig. \ref{fig:Figure7}(a), and consistent with previous reports that \LCO possesses short-range correlations well above $T_{N}$ \cite{ebisu1998,kawamata2004}. The appearance of features resembling gapped modes near 3 -- 3.5 meV in Fig. \ref{fig:Figure7}(a) is an artifact of the broad paramagnon scattering and the color scaling. The line cuts show no mode-like peak in this region and reducing the integration range along \textit{K} does not change this. 

These same slices and the zone center cut shown in the main panel of  Fig. \ref{fig:Figure7}(b) similarly indicate that the N = 2 peak vanishes at 20 K. However, in striking contrast, the intensity of the N = 3 peak is largely unchanged at 20 K, with only the overall background decreasing. Fitting this data (blue lines in Fig. \ref{fig:Figure7}(b)) shows that the $\Delta E$ value of the N = 3 mode changes slightly: $\Delta E_{N=3}$ = 4.00(2) meV and 4.25(4) meV at 5 K and 20 K, respectively. The N = 3 peak vanishes by $T$ = 150 K, indicating a finite regime of stability above $T_{N}$.

\begin{figure}[h]
 \includegraphics[width = 8.6 cm]{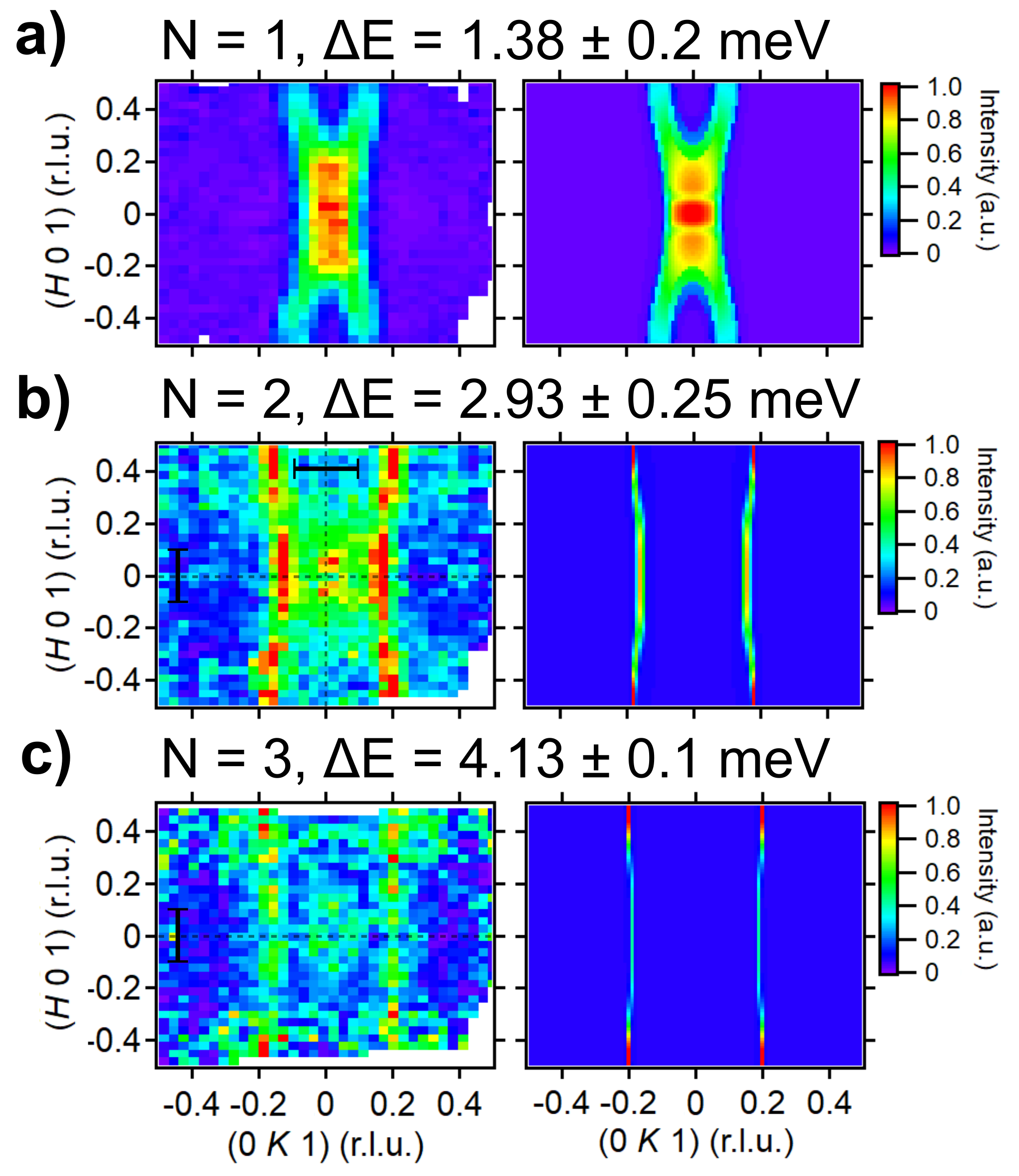}
 \caption{\label{fig:Figure8} Constant-$\Delta E$ slices in the ($H$ $K$ 1) plane from $E_{i}$ = 11 meV data  (left), and corresponding linear spin wave theory calculations using the exchange constants in Table \ref{tab:Table1} (right). No Al background has been subtracted. The extent of integration in $\Delta E$ is indicated at the top of each panel. The integration in the orthogonal direction in reciprocal space is $L$ = [0.95,1.05]. The calculated values have been convolved with the instrumental $\Delta E$-dependent resolution function. For N = 1, the $\va*{Q}$-resolution of the calculated data was set to the FWHM of the (0 0 1) magnetic Bragg peak. For N = 2 and 3, a larger $\va*{Q}$-resolution was needed to approximate the measured data. A constant value has been added to the calculated intensity to approximate an overall constant incoherent background. Within each plot, the intensity has been scaled by the maximum value. The black dashed lines and bars show the directions and integration regions, respectively, of the line cuts in Fig. \ref{fig:Figure9}. \textbf{a)} N = 1 mode. \textbf{b)} N = 2 mode \textbf{c)} N = 3 mode.}
 \end{figure}

\begin{figure*}[ht!]
 \includegraphics[width =17.4 cm]{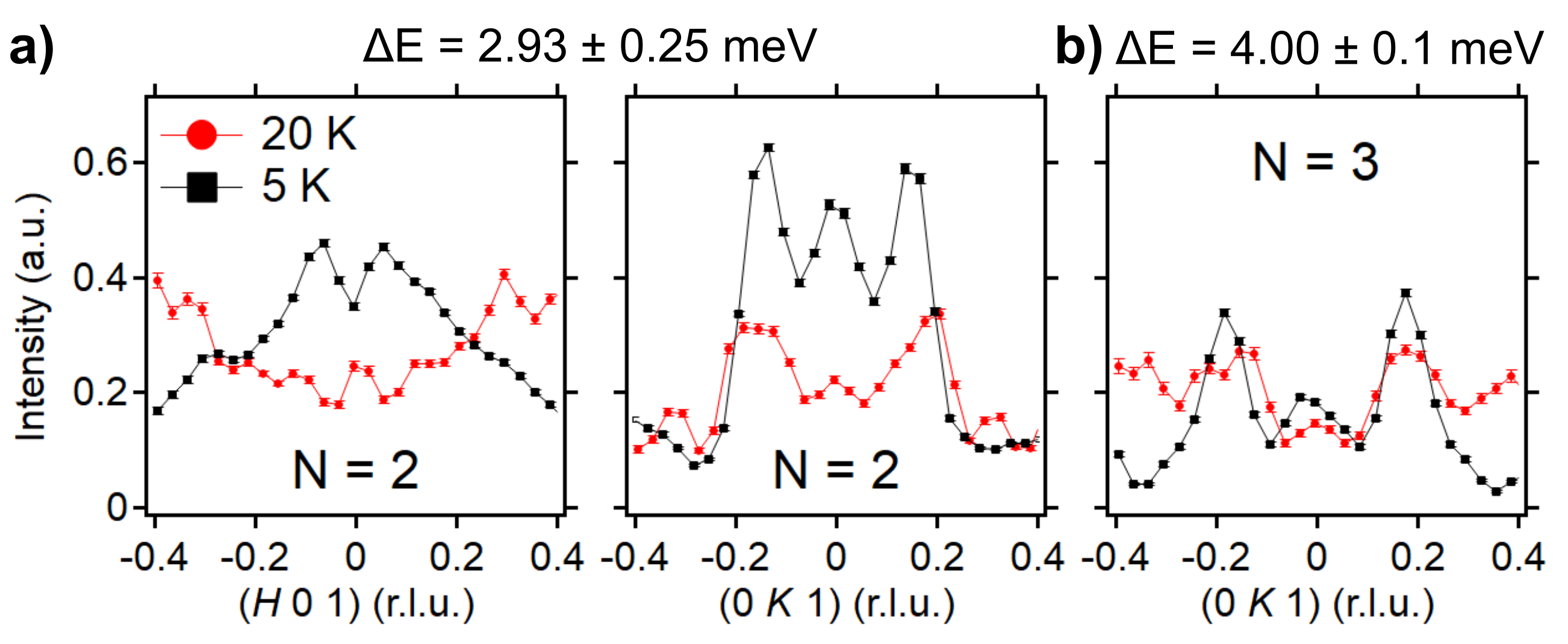}
 \caption{\label{fig:Figure9} Cuts through reciprocal space at constant $\Delta E$ at \textit{T} = 5 K and 20 K using $E_{i}$ = 11 meV data. The extent of integration in $\Delta E$, based on the fitted values from Fig. \ref{fig:Figure6}(c), is indicated at the top of the panels. All plots are on the same scale. \textbf{a)} N = 2 cuts along ($H$ 0 1) (left) and (0, $K$, 1) (right). The ($H$ 0 1) data was integrated in reciprocal space with the following bounds: $K$ = [-0.1, 0.1] and $L$ = [0.95, 1.05]. For the (0 $K$ 1) data the bounds are: $H$ = [-0.1, 0.1] and $L$ = [0.95, 1.05]. \textbf{b)} Cut along $K$ through the N = 3 excitation, with the following reciprocal space integration bounds: $H$ = [-0.1, 0.1] and $L$ = [0.95, 1.05]. }
 \end{figure*}

To study the momentum dependence of the N = 1, 2, and 3 modes in the ordered state, constant-$\Delta E$ slices through the ($H$ $K$ 1) plane are shown in the left panels of Fig. \ref{fig:Figure8}(a), (b), and (c). At higher $\Delta E$, the one-magnon, N = 1 dispersion presents as roughly parallel lines along the $H$-direction due to the much steeper dispersion along $K$, as shown in the simulated panels on the right hand side of Fig. \ref{fig:Figure8}. The N = 2 and N = 3 mode intensities emerge between these lines in the data, with the strongest scattering around the $\Gamma$-point.  Taking momentum cuts along the $K$-direction, Fig. \ref{fig:Figure9}(a), right, and Fig. \ref{fig:Figure9}(b) show the presence of a single peak around $K$ = 0. The stronger peaks at $K \approx \pm$ 0.2 on either side of the central peak come from the line-cuts crossing the N = 1 mode's dispersion. The persistence of signal in these regions above $T_{N}$ is, again, due to the strong paramagnon scattering. Line-cuts along the $H$-direction at $\Delta E_{N=2}$ reveal a two-peak structure around $H$ = 0, as shown in Fig. \ref{fig:Figure9}(a), left. A similar two-peak structure appears in line-cuts along the $L$-direction (not shown). Our current data set does not allow for a definitive statement on whether the same structure occurs along $H$ and $L$ at $\Delta E_{N=3}$; we aim to address this question in a future experiment. 

\section{Discussion \label{discussion}}
	
The expanded set of exchange parameters determined in fits to the one-magnon dispersion relation (summarized in Table \ref{tab:Table1}) remain consistent with the experimentally observed C-AFM structure when optimized within SpinW. These exchange parameters can be used to numerically test the conclusion from Lorenz \textit{et al.} \cite{lorenz2009} that the C-AFM state is stabilized primarily by $\tilde{J}_{131}$, with further stabilization from $D$. To do this, $\tilde{J}_{131}$ was set to 0 and $D$ was empirically modified to reproduce the measured spin gap, causing the optimized magnetic structure to transition into a spiral state. If $\tilde{J}_{131}$ is retained and the anisotropy is set to 0 (i.e., $D$ = 0 so that $J_{010}^{x} = J_{010}^{y} = J_{010}^{z}$) then the C-AFM structure results; however the predicted moment direction shifts to align along the $c$ axis. These results indicate that, within the expanded set of exchange parameters refined in our model, the C-AFM magnetic order remains stabilized by the $\tilde{J}_{131}$ interaction. 

We note here that later work from some of the authors of Lorenz \textit{et al.} \cite{lorenz2009} also explored the role of additional exchange parameters beyond the previously discussed minimal model \cite{lorenzThesis2011,kuzian2018}. The presented exchange model is similar to our final model, with small differences in the numerical values of the exchange constants; however, the need to include $J_{001}$ was not recognized nor was the relative importance of $J_{100}$ over $J_{110}$. Additionally, recent quantum chemistry calculations gave theoretical values for $J_{010}$ and $J_{020}$ ($\approx$ -215 K and 67 K, respectively) which match remarkably well to our experimentally extracted values \cite{matsuda2019}.

Our neutron scattering measurements reveal additional modes near the zone center outside of the one-magnon, N = 1, dispersion. Based on the $T$ = 5 K data shown in Fig. \ref{fig:Figure6}(c), the peak maximum of the first additional mode, $\Delta E_{N=2}$ = 2.924(5), is slightly above the value of 2$\Delta E_{N=1}$. This  is not due to an artificially decreased value of $\Delta E_{N=1}$ from this particular energy cut since the fitted value of  $\Delta E_{N=1}$ agrees within 0.02 meV with the calculated value from our exchange model. Variation of the finite momentum space integration produces a range of  $\Delta E_{N=2}$ values from 2.77 to 2.92 meV. This indicates that the observation of  $\Delta E_{N=2} > 2\Delta E_{N=1}$ is most likely a result of systematic error in our determination of the N = 2 mode energy. Note also that both the intrinsic energy resolution in this region and the utilized bin size are approximately equal to the difference $\Delta E_{N=2}$ $-$  2$\Delta E_{N=1}$.

In principle, two-magnon continuum scattering could be observable at the same energy scale and wave vector where the N = 2 mode is observed. However, the splitting seen within the N = 2 mode along $H$ [Fig. \ref{fig:Figure9}(a), left] and $L$ (not shown) is inconsistent with continuum scattering. The intensity of a two-magnon continuum is expected to be peaked at the magnetic zone center \cite{huberman2005} with structure possible near the zone boundary \cite{barnes2003}. Furthermore, we see no clear evidence of continuum-like features at higher energies. The appearance of a splitting around the zone center reflects additional interactions, consistent with the existence of a bound state. Such a splitting can potentially arise within a bound state when the potential that binds together the two interacting magnons (and which does not act upon the delocalized two-particle states of a continuum \cite{barnes2003}) possesses a $\va*{Q}$-dependence \cite{schneider1981}. Given the existence of significant interchain exchange interactions with components along $H$ and $L$, we ascribe this as the likely origin.

A result for the two-magnon binding energy in a 1D FM $J_1 - J_2$ \textit{XXZ}-Heisenberg model ($\alpha > 1/4$) was presented previously by Dmitriev and Krivnov \cite{dimitriev2009b}. Inputting our refined exchange parameters yields a small, theoretical binding energy of $\approx$ 0.03 meV, certainly within the resolution of our measurement. This calculation remains an estimate since it does not take into account the interchain interaction $\tilde{J}_{131}$, which is needed to stabilize the FM chains. The presence of such an interaction is expected to slightly reduce the binding energies of multimagnon bound states \cite{nishimoto2015}. Our observation of a peak in the spectrum within resolution of 2$\Delta E_{N=1}$ is qualitatively consistent with a small binding energy of this magnitude. Both N = 1 and N = 2 modes collapse upon warming above $T_N$, consistent with their formation via an interplay between the interchain coupling $\tilde{J}_{131}$ stabilizing the magnetic ground state and the Ising-like exchange anisotropy. 

The N = 3 mode is located at $\Delta E_{N=3}$ = 4.00(1) meV [see Fig. \ref{fig:Figure6}(c)] yielding 3$\Delta E_{N=1} - \Delta E_{N=3}$ = 0.08(1) meV, which exceeds the statistical error in the fitting procedure. However, the additional systematic error in the peak position introduced by the finite reciprocal space integration window means that this value likely falls within the resolution of our measurement. Generally, a three-magnon binding energy is expected to be larger than a two-magnon one, with the simplest picture being a factor of two increase. Given the estimation of the two-magnon binding energy, we would not expect to be able to resolve a three-magnon value. However, we are not currently aware of any theoretical calculation to which we can compare our results. While the value of $\Delta E_{N=3}$ is qualitatively consistent with a three-magnon bound state with a small binding energy, the persistence of the mode upon warming above $T_N$, where the one-magnon gap collapses, is anamolous and prevents a definitive identification. Clearly the stabilization mechanism for the N = 3 mode is distinct from that discussed previously for the N = 2, two-magnon bound state. The energy width of the N = 3 mode is 0.56(3) meV (Gaussian FWHM), relative to an instrumental resolution of 0.19 meV at $\Delta E=4.00$ meV, placing it well within an underdamped regime.

Models for three-magnon condensates have been proposed for the parameter space that \LCO occupies within $J_{1} - J_{2}$ ferromagnetic spin chain models \cite{kecke2007, nishimoto2011,nishimoto2015}. The survival of the N = 3 mode in a finite range above $T_N$ is potentially relevant to these models, though it would require an unconventional setting and the presence of a higher energy one-magnon mode not currently resolved in our measurements. Based on present calculations, these models also indicate that the relatively strong $\tilde{J}_{131}$ should move \LCO out of the regime of multipolar ground states \cite{nishimoto2011, nishimoto2015}. Utilization of our refined exchange parameters within these models may help resolve this question. Extension of recent numerical calculations of the dynamical structure factor at finite temperature to the specific case of \LCO might also be enlightening \cite{nayak2022}. On the experimental side, INS measurements exploring the polarization and dispersion of the N = 3 mode would be of use in identifying its origin. 

\section{Conclusions \label{conclusions}}

INS data exploring the magnetic excitations in \LCO were presented and used to refine the previously published exchange model. The new analysis confirms the importance of the $\tilde{J}_{131}$ interaction and determines the nature of further interchain interactions, both of which are consistent with the C-AFM ground state. Additional modes within the INS spectrum cannot be described in terms of the one-magnon dispersion, and we identify these as two- and, potentially, three-magnon bound states with very small binding energies. The persistence of the N = 3 mode in a finite range of $T > T_{N}$ is anomalous and motivates further investigation of \LCO as a platform for exchange anisotropy stabilized few-body bound states in a Heisenberg ferromagnetic chain.  

\FloatBarrier

\begin{acknowledgments}
We thank A. I. Kolesnikov for assistance with the inelastic neutron scattering experiment and acknowledge insightful discussions with L. Balents, S. Nishimoto, and C Agrapidis.  This work was supported by the US Department of Energy (DOE), Office of Basic Energy Sciences, Division of Materials Sciences and Engineering under Grant No. DE-SC0017752.  A portion of this research used resources at the Spallation Neutron Source, a DOE Office of Science User Facility operated by the Oak Ridge National Laboratory. This work made use of the MRL Shared Experimental Facilities which are supported by the MRSEC Program of the NSF under Award No. DMR 1720256, a member of the NSF-funded Materials Research Facilities Network. It also used facilities supported via the UC Santa Barbara NSF Quantum Foundry funded via the Q-AMASE-i program under award DMR-1906325. This work was additionally supported by the U.S. Department of Energy, Office of Science, Office of Workforce Development for Teachers and Scientists, Office of Science Graduate Student Research (SCGSR) program. The SCGSR program is administered by the Oak Ridge Institute for Science and Education for the DOE under contract number DE‐SC0014664. E.Z. recognizes financial support from JHU through the Sweeney Family Postdoctoral Fellowship.
\end{acknowledgments}

% Create the bibliography  using BibTeX:
%\bibliography{Bibliography}

\end{document}